\documentclass{appolb}
\usepackage{graphicx}
\usepackage[usenames]{color}

\begin{document}
\eqsec
\title{Continuous-time random walk model of relaxation of two-state systems}
\author{S.~I.~Denisov \thanks{denisov@sumdu.edu.ua}, Yu.~S.~Bystrik
\address{Sumy State University, Rimsky-Korsakov Street 2,
UA-40007 Sumy, Ukraine}
}
\maketitle
\begin{abstract}
Using the continuous-time random walk (CTRW) approach, we study the phenomenon of relaxation of two-state systems whose elements evolve according to a dichotomous process. Two characteristics of relaxation, the probability density function of the waiting times difference and the relaxation law, are of our particular interest. For systems characterized by Erlang distributions of waiting times, we consider different regimes of relaxation and show that, under certain conditions, the relaxation process can be non-monotonic. By studying the asymptotic behavior of the relaxation process, we demonstrate that heavy and superheavy tails of waiting time distributions correspond to slow and superslow relaxation, respectively.
\end{abstract}
\PACS{05.40.Fb, 02.50.Ey, 76.20.+q}

\section{Introduction}
\label{Intr}

The relaxation processes describing the transition of macroscopic systems from one equilibrium state to another are the subject of great interest. This is mainly because the characteristics of these processes contain important information about the mechanisms of relaxation. These relaxation processes are usually studied under the condition that a constant generalized force, which has been applied for a long time, is abruptly switched off (for details see Ref.~\cite{Datt}). In this case, one of the main characteristics of the relaxation process is an extensive thermodynamic variable conjugate to the generalized force. In particular, for dielectric relaxation (see, for example, \cite{Kao, BoSm, Ngai} and references therein) the pair, the generalized force and its conjugate variable, consists of the external electric field and the electric dipole moment of the relaxing system. Similarly, for magnetic relaxation \cite{Cow, Suhl, BMS} the corresponding pair consists of the external magnetic field and the magnetic moment of the system.

An important characteristic of any relaxation process is the relaxation law, i.e., the properly normalized (dimensionless) conjugate variable as a function of time. In the simplest case, when the rate of change of this function is proportional to its magnitude, the relaxation law is exponential. However, many systems exhibit anomalous, non-exponential, relaxation \cite{KRS}. For instance, the slow magnetic relaxation was discovered in systems of single-molecule \cite{SGCN, TLBG, SOPS} and single-chain \cite{CGLS, CMYC, SWG} magnets.

There is a wide class of systems whose relaxation properties, including anomalous ones, are completely characterized by the individual properties of their structural elements (such as single-molecule magnets, single-chain magnets, single-domain ferromagnetic particles, etc.). In particular, this happens when the state parameter of each element evolves according to a dichotomous random process (see below for a more detailed discussion). In this paper, we derive a number of analytical results describing the relaxation of these so-called two-state systems.

The paper is organized in the following way. In Sec.~\ref{DichMod}, we introduce the dichotomous model of relaxation of two-state systems, establish its connection with the CTRW model, find the Fourier-Laplace representation of the probability density function of the waiting times difference and the Laplace representation of the relaxation law. The laws of biased and unbiased relaxation of these systems characterized by Erlang distributions of waiting times are derived in Sec.~\ref{Relax}. Here we show that under certain conditions the relaxation process can be non-monotonic. In the same section, by studying the asymptotic behavior of relaxation processes in two-state systems, we demonstrate that heavy and superheavy tails of waiting time distributions are responsible for slow and superslow relaxation, respectively. Finally, our findings are summarized in Sec.~\ref{Concl}.

\section{Dichotomous model of relaxation}
\label{DichMod}

\subsection{Model description}

We consider a relaxing system as a set of identical objects, each of which can be in one of two states randomly changing in time. This model is widely used to describe the relaxation processes in physical systems whose structural elements are approximately characterized by two equilibrium states. Magnetic systems consisting of uniaxial single-domain ferromagnetic particles represent an important class of such systems. The magnetization of each particle has two equilibrium directions, but due to the thermal fluctuations, its instantaneous direction can be arbitrary. In this case, the magnetic relaxation is well described by using the stochastic Landau-Lifshitz or Landau-Lifshitz-Gilbert equation for magnetization dynamics and the corresponding Fokker-Planck equations for the probability density of the magnetization direction \cite{Suhl, DFT, CKW, CK}. This approach is quite general and may be applied to express the relaxation law through the system parameters. In particular, it has been used to approximately describe the features of magnetic relaxation arising from the dipole-dipole interaction between particles \cite{DT, DLT, Dej} and rotating external magnetic field \cite{DLHT, DSTH}.

However, the described approach is rather complicated technically. The main difficulty is the necessity to take into account all possible directions of the magnetization. But the role of the states (magnetization directions) that differ appreciably from the equilibrium ones decreases with decreasing temperature. Therefore, if the total probability of these states is small enough, the state parameter of particles, which describes the relaxation process, can be approximated by the dichotomous random process. This so-called dichotomous approximation will permit us to study in detail the relaxation processes in a whole class of two-state systems.

\begin{figure}[htb]
    \centerline{
    \includegraphics[totalheight=3cm]{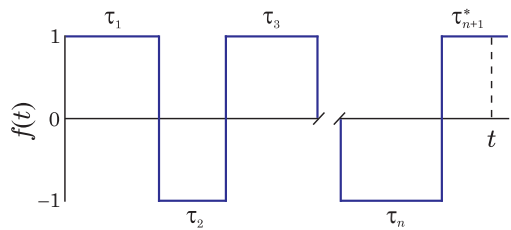}}
    \caption{\label{fig1} (Color online) A sample
    path of the dichotomous process $f(t)$ with
    even number of jumps in the interval $(0,t)$.}
\end{figure}

Within the dichotomous approximation, we associate the state parameter of each structural element of the system with the dichotomous (or telegraph) process $f(t)$, which takes the values $-1$ and $1$ and satisfies the initial condition $f(0)=1$, see Fig.~\ref{fig1}. This process is characterized by waiting (residence) times $\tau_{j}$ ($j = \overline{1,\infty}$), that is times between successive jumps of $f(t)$, which are assumed to be independent random variables. We also assume that the waiting times $\tau_{2j-1}$ in the up state, when $f(t) = 1$, and the waiting times $\tau_{2j}$ in the down state, when $f(t) = -1$, are distributed according to arbitrary probability densities $p^{+}(\tau)$ and $p^{-}(\tau)$, respectively.

The dichotomous process is the simplest random process that plays an important role in numerous applications. If the probability densities $p^{\pm}(\tau)$ are exponential, then many of the properties of this process can be determined by using the telegraph [if $p^{+}(\tau) = p^{-}(\tau)$] or generalized telegraph [if $p^{+}(\tau) \neq p^{-}(\tau)$] equations \cite{Gold, Kac, CoRa}. Since we are interested in studying the influence of different waiting time distributions on the character of relaxation of two-state systems, for this purpose it is more convenient to use the CTRW approach \cite{MoWe, AvHa, MeKl, MJCB}. In this approach, we are going to determine the probability density of the waiting times difference $\Delta_{t}$, that is the difference between the times that the dichotomous process $f(t)$ spends in the up and down states on the interval $(0,t)$, defined as
\begin{equation}
    \Delta_{t} = \int_{0}^{t}dt' f(t')
    \label{Delta_t},
\end{equation}
and study in detail the time dependence of the relaxation law
\begin{equation}
    \mu(t) = \mathrm{Pr} \{ f(t)\!=\!1 \} - \mathrm{Pr} \{ f(t)\!=\!-1 \},
    \label{Def_mu}
\end{equation}
where $\mathrm{Pr} \{ \cdot \}$ denotes the probability of the condition inside the braces.

In order to avoid any confusion, we emphasize that all sample paths of the dichotomous process $f(t)$ are assumed to start at $t=0$ and $f(0)=1$. If the sample paths started earlier, e.g., at $t=- \tau_{0}$ with $\tau_{0}>0$, then $f(t) =1$ as $t \in (-\tau_{0}, \tau_{1})$ and, due to ageing effects \cite{SBM}, the statistical properties of $\Delta_{t}$ and the relaxation function $\mu(t)$ could strongly depend on $\tau_{0}$. Therefore, to simplify the problem, we restrict ourselves to the case $\tau_{0}=0$.

\subsection{General results}

The probability density $P(\Delta,t)$ that the waiting times difference $\Delta_{t}$ equals $\Delta$ for a fixed time $t$ can be written as follows:
\begin{equation}
    P(\Delta,t) = \langle \delta(\Delta_{t} - \Delta) \rangle,
    \label{Def_P}
\end{equation}
where $\delta(\cdot)$ is the Dirac delta function and the angular brackets denote averaging over all sample paths of the dichotomous process $f(t)$. Since $\Delta_{t} \in (-t,t)$, one gets $P(\Delta,t) =0$ as $|\Delta|>t$, and the normalization condition for $P(\Delta,t)$ becomes $\int_{-t}^{t}d\Delta P(\Delta,t)=1$. Let us assume that the process $f(t)$ has exactly $n$ ($n=\overline{0, \infty}$) jumps in the interval $(0,t)$. Then, introducing the $n$-jump probability density
\begin{equation}
    P^{(n)}(\Delta,t) = \langle \delta(\Delta_{t} - \Delta) \rangle_{n}
    \label{Def_Pn}
\end{equation}
($ \langle \cdot \rangle_{n}$ denotes the average over these sample paths), we can represent $P(\Delta,t)$ in the form
\begin{equation}
    P(\Delta,t) = \sum_{n=0}^{\infty} P^{(n)}(\Delta,t).
    \label{P1}
\end{equation}

According to the definition (\ref{Def_Pn}), the probability that a given sample path of $f(t)$ has exactly $n$ jumps in the time interval $(0,t)$ is given by
\begin{equation}
    W^{(n)}(t) = \int_{-t}^{t} d\Delta P^{(n)}(\Delta,t),
    \label{Wn1}
\end{equation}
and the normalization condition for the probability density $P(\Delta,t)$ yields $\sum_{n=0}^{ \infty} W^{(n)}(t) =1$. Our next step is to express $W^{(n)}(t)$ and $P^{(n)}(\Delta,t)$ in terms of the waiting time probability densities $p^{\pm}(\tau)$. To this end, we first introduce the waiting times difference $\Delta_{t}^{(n)}$ for the sample paths with $n$ jumps in the interval $(0,t)$. It is obvious that $\Delta_{t}^{(0)}= t$, and if $n \geq 1$ then
\begin{equation}
    \Delta_{t}^{(n)} = \sum_{j=1}^n (-1)^{j-1}\tau_j+
    (-1)^n\tau_{n+1}^*,
    \label{Delta_t2}
\end{equation}
where
\begin{equation}
    \tau_{n+1}^* = t - \sum_{j=1}^n \tau_j \leq \tau_{n+1}.
    \label{tau*}
\end{equation}

The probability that a sample path of the dichotomous process $f(t)$ has no jumps on $(0,t)$, that is the probability that $\tau_{1} \geq t$, is written as
\begin{equation}
    W^{(0)}(t) = \int_{t}^{\infty}d\tau p^{+}(\tau).
    \label{W0}
\end{equation}
Let us now assume that the process $f(t)$ has $n \geq 1$ jumps in the interval $(0,t)$. If these jumps occur in the intervals $( \sum_{j=1}^{k} \tau_{j}, \sum_{j=1}^{k} \tau_{j} + d\tau_{k})$ with $k=\overline{1,n}$, then the probability $dW^{(n)}(t)$ of such process is given by
\begin{equation}
    dW^{(n)}(t) = \!\bigg( \prod_{j=1}^{n} d\tau_{j}
    p_j(\tau_{j}) \bigg)\! \int_{t - \sum_{j=1}^{n}
    \!\tau_{j}}^{\infty}\! d\tau p_{n+1}(\tau),
    \label{dWn}
\end{equation}
where $p_{j}(\tau) = p^{+}(\tau)$ or $p^{-}(\tau)$ if $j$ is odd or even, respectively. It is appropriate to recall here that Eq.~(\ref{dWn}) is obtained under conditions that (a) the jumps of $f(t)$ are independent events with the probabilities $p_j(\tau_{j}) d\tau_{j}$ and (b) the $(n+1)$st jump occurs outside the interval $(0,t)$. To avoid confusion, we note that the times $\tau_{j}$ in Eq.~(\ref{dWn}) (and in all probabilistic expressions below) are interpreted not as the random variables but as the variables of integration. Introducing the $n$-dimensional domain of integration $\Omega_n(t)$ defined by the condition $\sum_{j=1}^{n} \tau_j  \leq t$ and replacing the lower limit of integration in Eq.~(\ref{dWn}) by $\tau^{*}_{n+1}$, we can write the probability $W^{(n)}(t)$ in the form (see also Refs.~\cite{DKDH1})
\begin{equation}
    W^{(n)}(t) = \int_{\Omega_{n}(t)}\!
    \bigg( \prod_{j=1}^{n} d\tau_{j} p_j(\tau_{j}) \bigg)\!
    \int_{\tau^{*}_{n+1}}^{\infty}\! d\tau p_{n+1}(\tau).
    \label{Wn2}
\end{equation}

Finally, using the above results, we find
\begin{equation}
    P(\Delta,t) = W^{(0)}(t)\delta(t-\Delta) + \tilde{P}(\Delta,t),
    \label{P2}
\end{equation}
where $\tilde{P}(\Delta,t) = \sum_{n=1}^{\infty} P^{(n)}(\Delta,t)$ is the regular part of the probability density $P(\Delta,t)$ and
\begin{equation}
    P^{(n)}(\Delta,t) = \int_{\Omega_{n}(t)} \!
    \bigg( \prod_{j=1}^{n} d\tau_{j} p_j(\tau_{j}) \bigg)\!
    \int_{\tau^{*}_{n+1}}^{\infty}\! d\tau p_{n+1}(\tau)
    \delta \big(\Delta_{t}^{(n)} - \Delta \big).
    \label{Pn2}
\end{equation}
As it is seen from Eqs.~(\ref{P2}) and (\ref{Pn2}), $P(\Delta,t)$ depends on the waiting time densities $p^{+}(\tau)$ and $p^{-}(\tau)$ in a complicated manner. It seems reasonable therefore to find the connection between $P(\Delta,t)$ and $p^{\pm}(\tau)$ in Fourier-Laplace space. For this purpose, we first define the Fourier transform of a function $\varphi(\Delta)$ as $\mathcal{F} \{\varphi(\Delta)\} = \varphi_{k} = \int_{-\infty}^{\infty} d\Delta e^{ik\Delta} \varphi (\Delta)$ $(-\infty <k< \infty)$ and the Laplace transform of a function $\psi(t)$ as $\mathcal{L} \{\psi(t)\} = \psi_{s} = \int_{0}^{ \infty} dt e^{-st} \psi(t)$ $(\mathrm{Re}\, s>0)$. Then, taking the Fourier-Laplace transform of $P(\Delta,t)$ defined as $P_{ks} = \mathcal{L} \{ \mathcal{F} \{P(\Delta,t)\} \}$, from Eq.~(\ref{P2}) one obtains
\begin{equation}
   P_{ks} = \frac{1-p_{s-ik}^{+}}{s-ik} + \tilde{P}_{ks},
   \label{P_ks}
\end{equation}
where $\tilde{P}_{ks} = \sum_{n = 1}^{\infty} P_{ks}^{(n)}$ and, according to Eq.~(\ref{Pn2}),
\begin{eqnarray}
    P^{(n)}_{ks} \!& = &\! \int_{0}^{\infty}\! dt e^{-st}
    \bigg[\int_{\Omega_{n}(t)} \! \bigg( \prod_{j=1}^{n}
    d\tau_{j} p_j(\tau_{j}) \bigg) e^{ik\Delta_{t}^{(n)}}
    \nonumber\\[3pt]
    &&\! -  \int_{\Omega_{n+1}(t)} \! \bigg(\prod_{j=1}^{n+1}
    d\tau_{j} p_j(\tau_{j}) \bigg) e^{ik\Delta_{t}^{(n)}}
    \bigg]. \qquad \;
    \label{P_ks_n}
\end{eqnarray}

To calculate $P^{(n)}_{ks}$, we use the formula $\Delta_{t}^{(n)} = (-1)^{n}t - \sum_{j=1}^{n} [ (-1)^{n} + (-1)^{j} ]\tau_{j}$, which follows from Eqs.~(\ref{Delta_t2}) and (\ref{tau*}), and represent the inner integrals by the formula
\begin{equation}
    \int_{\Omega_{n}(t)} \! \prod_{j=1}^{n}
    d\tau_{j} p_{j}(\tau_{j}) = \int_{0}^{t}\! d\tau_{1}
    p_{1}(\tau_{1}) \int_{0}^{t-\tau_{1}}\! d\tau_{2}
    p_{2}(\tau_{2}) \ldots \int_{0}^{t- \sum_{j=1}^{n-1}\tau_{j}}\! d\tau_{n}
    p_{n}(\tau_{n}).
    \label{int}
\end{equation}
With these results, a straightforward integration in Eq.~(\ref{P_ks_n}) yields
\begin{equation}
    \begin{array}{cc}
    \displaystyle P_{ks}^{(2m-1)} = (p_{s-ik}^{+})^{m}
    (p_{s+ik}^{-})^{m-1}\, \frac{1-p_{s+ik}^{-}}{s+ik},
    \\ [10pt]
    \displaystyle P_{ks}^{(2m)} = (p_{s-ik}^{+}
    p_{s+ik}^{-})^m \,\frac{1-p_{s-ik}^{+}}{s-ik}
    \end{array}
    \label{P_ks_n2}
\end{equation}
$(m=\overline{1,\infty})$. Finally, using Eqs.~(\ref{P_ks_n2}) and the formula for the sum of an infinite geometric series, $\sum_{n=1}^{\infty} r^{n} = r/(1-r)$  ($|r|<1$), the Fourier-Laplace transform of $\tilde{P}(\Delta,t)$ can be written in the form
\begin{equation}
    \tilde{P}_{ks} = \frac{p_{s-ik}^{+}} {1-p_{s-ik}^{+}
    p_{s+ik}^{-}}\bigg(\frac{1 - p_{s+ik}^{-}}{s+ik} +
    \frac{1 - p_{s-ik}^{+}}{s-ik} p_{s+ik}^{-} \bigg).
    \label{tildeP_ks}
\end{equation}
Note that Eq.~(\ref{tildeP_ks}) with $p^{\pm}(\tau) = p(\tau)$ has been derived and used to determine the long-time behavior of $P(\Delta,t)$ in some particular cases \cite{GoLu}.

Equations (\ref{P_ks}) and (\ref{tildeP_ks}) represent the desired probability density $P(\Delta,t)$ in the Fourier-Laplace space. Because of the complex dependence of $P_{ks}$ on $k$ and $s$, the calculation of $P(\Delta,t)$ by taking the inverse Fourier-Laplace transform of $P_{ks}$ is possible only in exceptional cases. In particular, if $f(t)$ is the generalized telegraph process characterized by the exponential waiting time density functions
\begin{equation}
    p^{\pm}(\tau) = \lambda_{\pm} e^{-\lambda_{\pm} \tau}
    \label{p^pm}
\end{equation}
($\lambda_{\pm}>0$ are rate parameters), then
\begin{eqnarray}
    &P(\Delta,t)= e^{-\lambda_{+} t}
    \delta(t - \Delta) + \frac{\lambda_{+}}{2}
    \exp\!\Big(\! - \frac{\lambda_{+} +
    \lambda_{-}}{2}\,t - \frac{\lambda_{+} -
    \lambda_{-}}{2}\,\Delta \Big)&
    \nonumber\\[4pt]
    &\times \Big[ I_{0}\Big(\sqrt{\lambda_{+}
    \lambda_{-}(t^2 - \Delta^2)}\,\Big) +
    \sqrt{\frac{\lambda_{-}}{\lambda_{+}}}
    \sqrt{\frac{t+\Delta}{t-\Delta}}\,I_{1}
    \Big(\sqrt{\lambda_{+}\lambda_{-}(t^2 -
    \Delta^2)}\,\Big) \Big],&\quad
    \label{mathcal_P2}
\end{eqnarray}
where $I_{0}(\cdot)$ and $I_{1}(\cdot)$ are the modified Bessel functions of the first kind of order zero and one, respectively. It should be noted that this result was previously obtained using other methods (see Refs.~\cite{CoRa, BNO, Uch}).

We are also concerned with the time dependence of the relaxation function $\mu(t)$. According to the definition (\ref{Def_mu}), this function satisfies the conditions $\mu(0)=1$ and $\mu(t) \in [-1,1]$. Then, since $\langle \Delta_{t}  \rangle = \int_{0}^{t} dt' \langle f(t') \rangle$ and $\langle f(t) \rangle = \mu(t)$, it can be represented in the form $\mu(t) = d \langle \Delta_{t} \rangle/dt$, which is useful if $\langle \Delta_{t} \rangle$ as a function of time is known. But to study the time dependence of $\mu(t)$ on waiting time distributions, it is more convenient to use the Laplace representation of $\mu(t)$. To derive it, we first express the relaxation function in terms of the probabilities $W^{(n)}(t)$:
\begin{equation}
    \mu(t) = W^{(0)}(t) + \sum_{m=1}^{\infty}
    \big[ W^{(2m)}(t) - W^{(2m-1)}(t) \big].
    \label{mu}
\end{equation}
Then, applying the Laplace transform to Eq.~(\ref{mu}) and taking into account that $W^{(0)}_{s} = (1 - p^{+}_{s})/s$ and $W^{(n)}(t) = P^{(n)}_k(t)\big|_{k=0}$, we obtain
\begin{equation}
    \mu_{s}= \frac{1-p^{+}_s}{s} + \sum_{m=1}^{\infty}
    \big( P^{(2m)}_{ks} - P^{(2m-1)}_{ks}\big)\big|_{k=0}.
    \label{mu_s2}
\end{equation}
Finally, using Eqs.~(\ref{P_ks_n2}) and performing summation over $m$, Eq.~(\ref{mu_s2}) can be reduced to
\begin{equation}
    \mu_{s}= \frac{1-2p^{+}_{s} + p^{+}_{s} p^{-}_{s}}
    {s(1- p^{+}_{s} p^{-}_{s})}.
    \label{mu_s3}
\end{equation}

This result, which holds for arbitrary waiting time distributions, is our main tool for studying the relaxation of two-state systems. Before we proceed to the analysis of relaxation laws, it is worthwhile to note that in the case of biased relaxation [when $p^{+}(\tau) \neq p^{-}(\tau)$] Eq.~(\ref{mu_s3}) is equivalent to the integral equation
\begin{eqnarray}
    \mu(t) \!&-&\! \int_0^t d\tau \mu(\tau)\int_0^{t-\tau} d\tau'p^+(\tau')p^-(t-\tau-\tau')
    \nonumber\\[4pt]
    \!&=&\! 1 - 2\int_{0}^{t}d\tau p^{+}(\tau)
    + \int_0^{t}d\tau p^+(\tau) \int_{0}^{t-\tau}d\tau'
    p^{-}(\tau').
    \label{eq_mu}
\end{eqnarray}
In the case of unbiased relaxation [when $p^{\pm}(\tau) = p(\tau)$] Eqs.~(\ref{mu_s3}) and (\ref{eq_mu}) are simplified to
\begin{equation}
    \mu_{s} = \frac{1-p_{s}}{s(1+p_{s})}
    \label{mu_s4}
\end{equation}
and
\begin{equation}
    \mu(t) + \int_{0}^{t} d\tau \mu(\tau) p(t-\tau)
    = 1 - \int_{0}^{t}d\tau p(\tau),
    \label{eq_mu2}
\end{equation}
respectively. Equations (\ref{eq_mu}) and (\ref{eq_mu2}) show that, in general, the memory effects in two-state systems play an important role for both biased and unbiased regimes of relaxation.

\section{Relaxation laws}
\label{Relax}

\subsection{Exact results}

Here we consider the two-state systems characterized by Erlang distributions of waiting times, whose probability density functions are given by
\begin{equation}
    p^{\pm}(\tau) = \frac{\lambda_{\pm}^{k}\tau^{k-1}}
    {(k-1)!}\, e^{-\lambda_{\pm} \tau},
    \label{Erlang}
\end{equation}
where $k = \overline{1,\infty}$ is the shape parameter. Since many of the properties of such systems have already been studied \cite{DiCr}, we are only concerned with the relaxation function $\mu(t)$. The conditions $\lambda_{+} \neq \lambda_{-}$ and $\lambda_{+} = \lambda_{-} = \lambda$ correspond to the biased and unbiased relaxation, respectively.

\subsubsection{Biased relaxation at $k=1$}

At $k=1$ the Erlang density functions (\ref{Erlang}) become exponential. According to Eq.~(\ref{p^pm}), in this case $p^{\pm}_{s} = \lambda_{\pm}/ (\lambda_{\pm} + s)$ and Eq.~(\ref{mu_s3}) yields
\begin{equation}
    \mu_{s}= \frac{s - \lambda_{+} + \lambda_{-}}
    {s(s + \lambda_{+} + \lambda_{-})}.
    \label{mu_s_exp}
\end{equation}
By applying to Eq.~(\ref{mu_s_exp}) the inverse Laplace transform [see Ref.~\cite{Erd}, Eq.~5.2(5)], we make sure that the relaxation function is purely exponential:
\begin{equation}
    \mu(t) = \frac{\lambda_{-} - \lambda_{+}} {\lambda_{+}
    + \lambda_{-}} + \frac{2\lambda_{+}} {\lambda_{+}
    +\lambda_{-}}e^{-(\lambda_{+} + \lambda_{-})t}.
    \label{mu2}
\end{equation}
It can be easily verified that this function satisfies the differential equation
\begin{equation}
    \frac{d}{dt}\mu(t) + (\lambda_{+} + \lambda_{-})
    \mu(t) + \lambda_{+} - \lambda_{-} = 0
    \label{eq_mu3}
\end{equation}
($\mu(0)=1$), which also follows from Eq.~(\ref{eq_mu}). Thus, there are no memory effects in this case.

\subsubsection{Biased relaxation at $k=2$}

At $k=2$ from Eqs.~(\ref{Erlang}) and (\ref{mu_s3}) one can get $p^{\pm}_{s} = \lambda_{\pm}^{2}/ (s + \lambda_{\pm})^{2}$  and
\begin{equation}
    \mu_{s}= \frac{s^{3} + 4as^{2} + 4(\nu^{2} + ab)s +
    4\nu^{2} b}{s(s + s_{+})(s + s_{-})(s + 2a)},
    \label{mu_s5}
\end{equation}
where $a = (\lambda_{+} + \lambda_{-})/2$, $b = (\lambda_{-} - \lambda_{+})/2$, $\nu = \sqrt{\lambda_{+}\lambda_{-}}$, and $s_{\pm} = a \pm \sqrt{a^{2} - 2\nu^{2}}$. Taking the inverse Laplace transform of Eq.~(\ref{mu_s5}) [see Ref.~\cite{Erd}, Eq.~5.2(19)], for a given class of two-state systems we obtain
\begin{eqnarray}
    \mu(t) \!\!&=&\!\! \frac{\lambda_{-} - \lambda_{+}}
    {\lambda_{+} + \lambda_{-}}\bigg(1 + \frac{\lambda_{+}}
    {\lambda_{-}} e^{-(\lambda_{+} + \lambda_{-})t}\bigg)
    + \frac{\lambda_{+}} {\lambda_{-}}\bigg[
    \cosh{(\sqrt{a^{2} - 2\nu^{2}}\,t)}
    \nonumber\\[6pt]
    &&\!\! - (\lambda_{+} - 3\lambda_{-}) \frac{\sinh{
    (\sqrt{a^{2} - 2\nu^{2}} \,t)}}{2\sqrt{a^{2} -
    2\nu^{2}}} \bigg]e^{-(\lambda_{+} + \lambda_{-})t/2}.
    \label{mu3}
\end{eqnarray}
Although the relaxation functions (\ref{mu2}) and (\ref{mu3}) have the same limiting value $\mu(\infty) = (\lambda_{-} - \lambda_{+})/(\lambda_{+} + \lambda_{-})$, their behavior at finite times is quite different. In particular, at small times the function $1 - \mu(t)$ is proportional to $t$ and $t^{2}$, respectively. Moreover, in contrast to (\ref{mu2}), the relaxation function (\ref{mu3}) at $a^{2} - 2\nu^{2} > 0$ is characterized by three relaxation times, the largest of which is $1/s_{-}$.

The most important qualitative difference between the relaxation laws (\ref{mu2}) and (\ref{mu3}) occurs if $a^{2} - 2\nu^{2} < 0$, that is if the parameters $\lambda_{\pm}$ satisfy the conditions
\begin{equation}
    3 - 2\sqrt{2} < \frac{\lambda_{+}}{\lambda_{-}}
    < 3 + 2\sqrt{2}.
    \label{cond}
\end{equation}
Since in this case $\sqrt{a^{2} - 2\nu^{2}} = i \sqrt{2\nu^{2} - a^{2}}$ with $2\nu^{2} - a^{2}>0$, the hyperbolic functions in Eq.~(\ref{mu3}) should be replaced by the trigonometric ones. Therefore, introducing the period of these functions as $T = 2\pi/\sqrt{2\nu^{2} - a^{2}}$ or
\begin{equation}
    T =  \frac{4\pi}{\sqrt{6\lambda_{+}\lambda_{-}
    - \lambda_{+}^{2} - \lambda_{-}^{2}}},
    \label{T}
\end{equation}
Eq.~(\ref{mu3}) can be rewritten in the form
\begin{eqnarray}
    \mu(t) \!\!&=&\!\! \frac{\lambda_{-} - \lambda_{+}}
    {\lambda_{+} + \lambda_{-}}\bigg(1 + \frac{\lambda_{+}}
    {\lambda_{-}} e^{-(\lambda_{+} + \lambda_{-})t}\bigg)
    + \frac{\lambda_{+}} {\lambda_{-}}\bigg[
    \cos \! \bigg( \frac{2\pi t}{T} \bigg)
    \nonumber\\[6pt]
    &&\!\! - (\lambda_{+} - 3\lambda_{-})\frac{T}{4\pi}
    \sin \! \bigg( \frac{2\pi t}{T}\bigg) \bigg]
    e^{-(\lambda_{+} + \lambda_{-})t/2}.
    \label{mu4}
\end{eqnarray}

\begin{figure}[htb]
    \centerline{
    \includegraphics[totalheight=4.5cm]{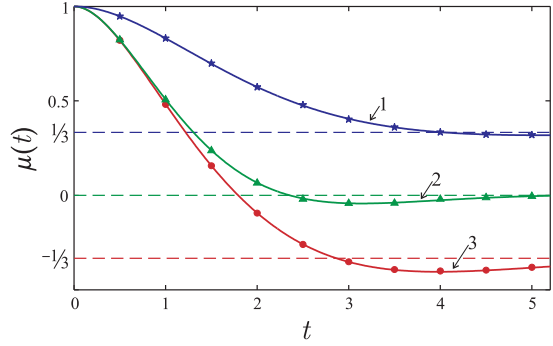}}
    \caption{\label{fig2} (Color online) The relaxation functions
    in the case of Erlang distributions of waiting times for $k=2$,
    $\lambda_{+} = 0.5,\, \lambda_{-} = 1$ (1), $\lambda_{+}
    = \lambda_{-} = 1$ (2), and $\lambda_{+} = 1,\, \lambda_{-}
    = 0.5$ (3). The solid lines represent the theoretical result
    (\ref{mu4}) and the symbols show the simulation results for
    $\langle f(t) \rangle$. The period $T$ of trigonometric
    functions in Eq.~(\ref{mu4}) equals $8\pi/\sqrt{7}$ and
    $2\pi$ for the cases (1), (3) and (2), respectively.}
\end{figure}

According to this formula, the function $\mu(t)$ tends to the limiting value $\mu(\infty)$ in an oscillating manner (if the conditions (\ref{cond}) hold). This is a remarkable and somewhat unexpected result because the waiting time densities do not contain periodic functions. In view of our previous discussion, such behavior of $\mu(t)$ can be interpreted as the emergence of memory effects. The non-monotonic dependence of $\mu(t)$ on $t$ obtained from Eq.~(\ref{mu4}) is illustrated in Fig.~\ref{fig2} for different values of the rate parameters $\lambda_{\pm}$.

\subsubsection{Unbiased relaxation at $k \geq 2$}

In this case $p_{s} = \lambda^{k}/ (s + \lambda)^{k}$ and, according to Eq.~(\ref{mu_s4}), we obtain
\begin{equation}
    \mu_{s}= \frac{1 - (s + \lambda)^{k}/ \lambda^{k}}
    {[1 - (s + \lambda)/ \lambda][1 + (s + \lambda)^{k}/
    \lambda^{k}]}.
    \label{mu_s6}
\end{equation}
The last result shows that $\mu(t) = e^{-\lambda t} g(\lambda t)$, where the function $g(t)$ is defined by its Laplace transform
\begin{equation}
    g_{s}= \frac{1 - s^{k}}{(1 - s)(1 + s^{k})}.
    \label{g_s1}
\end{equation}
Using the relations $(1 - s^{k})/(1-s) = \sum_{j=1}^{k}s^{j-1}$ and $1 + s^{k} = \prod_{j=1}^{k}(s - a_{j})$ with $a_{j} = e^{i\pi (2j - 1)/k}$ being the solution of the equation $1 + s^{k} = 0$, we can rewrite the above formula as
\begin{equation}
    g_{s} = \frac{\sum_{j=1}^{k}s^{j-1}}
    {\prod_{j=1}^{k}(s - a_{j})}.
    \label{g_s2}
\end{equation}
Then, taking the inverse Laplace transform of Eq.~(\ref{g_s2}) [see again Ref.~\cite{Erd}, Eq.~5.2(19)], we find
\begin{equation}
    g(t) = \sum_{l=1}^{k}\frac{\sum_{j=1}^{k}a^{j-1}_{l}}
    {\prod_{j=1}'^{k}(a_{l} - a_{j})}\, e^{a_{l}t},
    \label{g(t)}
\end{equation}
where the prime on the product means that $j \neq l$. This result, together with the fact that $\sum_{j=1}^{k}a^{j-1}_{l} = 2/(1 - a_{l})$, yields
\begin{equation}
    \mu(t) = 2\sum_{l=1}^{k}\frac{e^{-(1 - a_{l})
    \lambda t}} {(1 - a_{l})\prod_{j=1}'^{k}(a_{l}
    - a_{j})}.
    \label{mu5}
\end{equation}
Finally, since $\prod_{j=1}'^{k} (a_{l}-a_{j}) = k a_{l}^{k-1} = -k/a_{l}$, the relaxation function (\ref{mu5}) can be represented in the form
\begin{equation}
    \mu(t) = \frac{\theta_k}{k}e^{-2\lambda t}+
    \frac{2}{k}\sum_{l=1}^{[k/2]} \left[\cos(
    \beta_l \lambda t)+ \frac{\beta_l}{1 - \alpha_l}
    \sin(\beta_l \lambda t)\right]\! e^{-(1-\alpha_l)
    \lambda t}.
    \label{mu_trig}
\end{equation}
Here, $\theta_k=0$ or $1$ if $k$ is even or odd, respectively, $[k/2]$ is the integer part of $k/2$, and
\begin{equation}
    \alpha_l = \cos\! \left(\frac{2\pi l
    -\pi}{k} \right)\!, \quad
    \beta_l = \sin\! \left(\frac{2\pi l -\pi}{k}
    \right)\!.
    \label{par_mu_trig}
\end{equation}

Thus, according to Eq.~(\ref{mu_trig}), the relaxation to the final state [$\mu(\infty) = 0$] also occurs in an oscillating way. But, in contrast to the relaxation function (\ref{mu4}) that oscillates with a single period $T$, in this case the oscillating part of $\mu(t)$ is, in general, characterized by a few periods $T_{l} = 2\pi /(\beta_{l} \lambda)$ (if $k \geq 5$).

\subsection{Asymptotic results}

Next we study the long-time behavior of the relaxation function in two types of systems. One of them is characterized by heavy tails of waiting time distributions, and the other by superheavy tails. Our aim here is to find universal asymptotic laws for unbiased relaxation, when $p^{\pm}(\tau) = p(\tau)$.

\subsubsection{Heavy-tailed $p(\tau)$}

A class of heavy-tailed probability density functions of waiting times is defined by the asymptotic behavior
\begin{equation}
    p(\tau) \sim \frac{q}{\tau^{1 + \alpha}}
    \label{as_p1}
\end{equation}
$(\tau \to \infty)$, where $q$ is a positive parameter and the tail index $\alpha$ satisfies the condition $\alpha \in (0,2]$. Since the long-time behavior of $\mu(t)$ is related to the behavior of $\mu_{s}$ when the real parameter $s$ tends to zero \cite{Fel, Hug}, we need to find $p_{s}$ as $s \to 0$. Using the asymptotic  formula (\ref{as_p1}), it is not difficult to show (see, e.g., Ref.~\cite{DDK}) that
\begin{equation}
    1 - p_{s} \sim \left\{\!\! \begin{array}{ll}
    q\frac{\Gamma(1-\alpha)}{\alpha}s^{\alpha},
    & \alpha \in (0,1)
    \\ [6pt]
    qs\ln\frac{1}{s},
    & \alpha = 1
    \\ [6pt]
    \bar{\tau}s - q\frac{\Gamma(2-\alpha)}
    {\alpha(\alpha-1)}s^{\alpha},
    & \alpha \in (1,2)
    \\ [6pt]
    \bar{\tau}s - \frac{q}{2}s^{2}\ln \frac{1}{s},
    & \alpha = 2
    \end{array}
    \right.
    \label{1-ps1}
\end{equation}
($s \to 0$), where $\Gamma(\cdot)$ is the gamma function and $\bar{\tau} = \int_{0}^{\infty} d\tau \tau p(\tau)$. Based on this result, the asymptotic behavior of $\mu(t)$ can be obtained from the Tauberian theorem for Laplace transforms. It states \cite{Fel, Hug} that if a function $h(t)$ is ultimately monotone and
\begin{equation}
    h_{s} \sim \frac{1}{s^{\rho}} L\! \left( \frac{1}{s} \right)
    \label{h_s}
\end{equation}
as $s \to 0$ then
\begin{equation}
    h(t) \sim \frac{1}{\Gamma(\rho)}\, t^{\rho -1}L(t)
    \label{h(t)}
\end{equation}
as $t \to \infty$. Here, $\rho >0$ and $L(t)$ is a positive function slowly varying at infinity, i.e., function for which the condition $L(\sigma t) \sim L(t)$ $(t \to \infty)$ holds for all $\sigma >0$.

If $\alpha \in (0,1)$ then, according to Eqs.~(\ref{mu_s4}) and (\ref{1-ps1}), $\mu_{s} \sim q\Gamma(1-\alpha)/(2\alpha s^{1-\alpha})$ ($s \to 0$) and, associating $h(t)$ with $\mu(t)$, directly from the Tauberian theorem one gets
\begin{equation}
    \mu(t) \sim \frac{q}{2\alpha}\, t^{-\alpha}
    \label{mu6}
\end{equation}
as $t \to \infty$. Since at $\alpha = 1$ and $s \to 0$ the condition $\mu_{s} \sim (q/2) \ln (1/s)$ occurs, the Tauberian theorem is not applicable to $\mu(t)$. However, it can be applied to the function $h(t) = \int_{0}^{t} dt'\mu(t')$ whose Laplace transform is given by $h_{s} = \mu_{s}/s$. This yields $h(t) \sim (q/2) \ln t$ as $t \to \infty$ and, because $\mu(t) = dh(t)/dt$, one can make sure that the case $\alpha = 1$ is described by the asymptotic formula (\ref{mu6}) as well. Similarly, introducing the auxiliary function $h(t) = \bar{\tau}/2 - \int_{0}^{t} dt'\mu(t')$ and using the relation $\mu(t) = -dh(t)/dt$, it is not difficult to verify that the above formula holds also for $\alpha \in (1,2]$. Thus, any two-state system with a heavy-tailed distribution of waiting times exhibits slow (power-law) relaxation (\ref{mu6}).

\subsubsection{Superheavy-tailed $p(\tau)$}

Let us finally consider a class of superheavy-tailed probability density functions of waiting times defined by the asymptotic behavior
\begin{equation}
    p(\tau) \sim \frac{1}{\tau}\, l(\tau)
    \label{as_p2}
\end{equation}
$(\tau \to \infty)$. Here, $l(\tau)$ is a slowly varying function which, due to the normalization condition $\int_{0}^{\infty} d\tau p(\tau) = 1$, tends to zero in such a way that $l(\tau) = o(1/\ln \tau)$ as $\tau \to \infty$. This class of probability densities has recently been used to study the phenomenon of superslow diffusion within the CTRW formalism \cite{DeK1, DeK2, DYBK, DBK}.

To find the long-time behavior of the relaxation function $\mu(t)$ in such two-state systems, we first introduce the exceedance probability $V(t) = \int_{t}^{\infty} d\tau p(\tau)$. Its important feature, which follows directly from Eq.~(\ref{as_p2}), is that it tends to zero at $t \to \infty$ as a slowly varying function. This fact enables us to determine a universal asymptotic formula for $\mu(t)$. Indeed, since $1 - p_{s} = sV_{s}$ and $V_{s} \sim (1/s)V(1/s)$ $(s \to 0)$, Eq.~(\ref{mu_s4}) yields $\mu_{s} \sim (1/2s)V(1/s)$. From this, using the Tauberian theorem, one obtains
\begin{equation}
    \mu(t) \sim \frac{1}{2}\, V(t)
    \label{mu7}
\end{equation}
as $t \to \infty$. Since the condition $\lim_{t \to \infty} t^{\rho}V(t) = \infty$ holds for all $\rho>0$ \cite{BGT}, $\mu(t)$ decreases slower than any negative power of time. It is therefore the relaxation process, whose long-time behavior is described by Eq.~(\ref{mu7}), can be called superslow. In particular, if
\begin{equation}
    p(\tau) = \frac{\ln c}{(c + \tau)\ln^{2}(c + \tau)}
    \label{super_p}
\end{equation}
$(c>1)$ then the exceedance probability equals $V(t) = \ln c/\ln(c + t)$ and Eq.~(\ref{mu7}) gives
\begin{equation}
    \mu(t) \sim \frac{\ln c}{2\ln t}.
    \label{mu8}
\end{equation}

\begin{figure}[htb]
    \centerline{
    \includegraphics[totalheight=5.5cm]{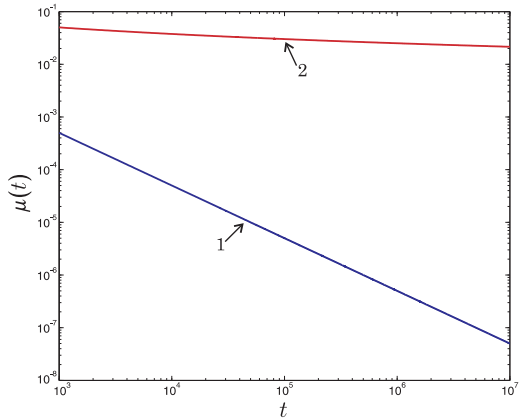}}
    \caption{\label{fig3} (Color online) Asymptotic behavior
    of the relaxation function $\mu(t)$ in regimes of slow 
    (1) and superslow (2) relaxation. The line 1 represents 
    the asymptotic result (\ref{mu6}) with $q=1$ and 
    $\alpha =1$, and the line 2 represents the asymptotic 
    result (\ref{mu8}) with $c=2$.}
\end{figure}

The asymptotic behavior of the relaxation function $\mu(t)$ in the cases of heavy and superheavy tails of the waiting time distribution $p(\tau)$ is illustrated in Fig.~\ref{fig3}. It shows that, because $V(t)$ is a slowly varying function, the function $\mu(t)$ in the regime of superslow relaxation tends to zero much slower than that in the regime of slow relaxation.

\section{Conclusions}
\label{Concl}

Using the continuous-time random walk approach, we have studied the phenomenon of relaxation in a class of two-state systems, whose structural elements evolve according to the dichotomous process. Our interest has been focused on the probability density function of the waiting times difference and on the relaxation law. Assuming that the distributions of waiting times in the up and down states of the dichotomous process are arbitrary, we have found the Fourier-Laplace representation of this density function and the Laplace representation of the relaxation function. These representations have been used to determine the density function in the case of exponential distributions of waiting times, to derive the integral equations describing the biased and unbiased relaxation processes, and to calculate the relaxation laws in some special cases. In particular, we have considered the Erlang distributions of waiting times and two classes of distributions characterized by heavy and superheavy tails, respectively. It has been shown that, depending on the waiting time distributions, the two-state systems can exhibit a wide variety of relaxation regimes including exponential, oscillatory, slow and superslow ones.

\section*{Acknowledgments}

We are grateful to the Ministry of Education and Science of Ukraine for financial support under Grant No.~0112U001383.


\begin{thebibliography} {99}

\bibitem{Datt} S.~Dattagupta, \textit{Relaxation Phenomena in Condensed Matter Physics} (Academic Press, Orlando, 1987).

\bibitem{Kao} K.~C.~Kao, \textit{Dielectric Phenomena in Solids} (Elsevier Academic Press, San Diego, 2004).
\bibitem{BoSm} R.~H.~Boyd and G.~D.~Smith, \textit{Polymer Dynamics and  Relaxation} (Cambridge University Press, Orlando, 2007).
\bibitem{Ngai} K.~L.~Ngai, \textit{Relaxation and Diffusion in Complex Systems} (Springer, New York, 2011).

\bibitem{Cow} B.~Cowan, \textit{Nuclear Magnetic Resonance and Relaxation} (Cambridge University Press, Cambridge, 2005).
\bibitem{Suhl} H.~Suhl, \textit{Relaxation Processes in Micromagnetics} (Oxford University Press, Oxford, 2007).
\bibitem{BMS} G.~Bertotti, I.~Mayergoyz, and C.~Serpico, \textit{Nonlinear Magnetization Dynamics in Nanosystems} (Elsevier, Oxford, 2009).

\bibitem{KRS} R.~Klages, G.~Radons, and I.~M.~Sokolov, eds.,     \textit{Anomalous Transport: Foundations and Applications} (Wiley-VCH, Berlin, 2008).

\bibitem{SGCN} R.~Sessoli, D.~Gatteschi, A.~Caneschi, and M.~A.~Novak, Nature \textbf{365}, 141 (1993).
\bibitem{TLBG} L.~Thomas, F.~Lionti, R.~Ballou, D.~Gatteschi, R.~Sessoli, and B.~Barbara, Nature \textbf{383}, 145 (1996).
\bibitem{SOPS} C.~Sangregorio, T.~Ohm, C.~Paulsen, R.~Sessoli, and D.~Gatteschi,  Phys.\ Rev.\ Lett. \textbf{78}, 4645 (1997).

\bibitem{CGLS} A.~Caneschi, D.~Gatteschi, N.~Lalioti, C.~Sangregorio, R.~Sessoli, G.~Venturi, A.~Vindigni, A.~Rettori, M.~G.~Pini, and M.~A.~Novak, Angew.\ Chem.\ Int.\ Ed.\ \textbf{40}, 1760 (2001).
\bibitem{CMYC} R.~Cl\'{e}rac, H.~Miyasaka, M.~Yamashita, and C.~Coulon, J.\ Am.\ Chem.\ Soc.\ \textbf{124}, 12837 (2002).
\bibitem{SWG} H.-L.~Sun, Z.-M.~Wang, and S.~Gao, Coord.\ Chem.\ Rev.\ \textbf{254}, 1081 (2010).

\bibitem{DFT} J.~L.~Dormann, D.~Fiorani, and E.~Tronc, Adv.\ Chem.\ Phys.\  \textbf{98}, 283 (1997).
\bibitem{CKW} W.~T.~Coffey, Yu.~P.~Kalmykov, and J.~T.~Waldron, \textit{The Langevin Equation}, 2nd ed.\ (World Scientific,  Singapore, 2004).
\bibitem{CK} W.~T.~Coffey and Yu.~P.~Kalmykov, J.\ Appl.\ Phys.\ \textbf{112}, 121301 (2012).

\bibitem{DT} S.~I.~Denisov and K.~N.~Trohidou,  Phys.\ Rev.\ B  \textbf{64}, 184433 (2001).
\bibitem{DLT} S.~I.~Denisov, T.~V.~Lyutyy, and K.~N.~Trohidou,  Phys.\ Rev.\ B  \textbf{67}, 014411 (2003).
\bibitem{Dej} P.~M.~D\'{e}jardin, J.\ Appl.\ Phys.\ \textbf{110}, 113921 (2011).

\bibitem{DLHT} S.~I.~Denisov, T.~V.~Lyutyy, P.~H\"{a}nggi, and K.~N.~Trohidou,  Phys.\ Rev.\ B  \textbf{74}, 104406 (2006).
\bibitem{DSTH} S.~I.~Denisov, K.~Sakmann, P.~Talkner, and P.~H\"{a}nggi, Phys.\ Rev.\ B  \textbf{75}, 184432 (2007).

\bibitem{Gold} S.~Goldstein, Quart.\ J.\ Mech.\ Appl.\ Math.\ \textbf{4}, 129 (1951).
\bibitem{Kac} M.~Kac, Rocky Mount.\ J.\ Math.\ \textbf{4}, 497 (1974).
\bibitem{CoRa} A.~D.~Kolesnik and N.~Ratanov, \textit{Telegraph Processes and Option Pricing} (Springer, Heidelberg, 2013).

\bibitem{MoWe}
    E.~W.~Montroll and G.~H.~Weiss, J.\ Math.\ Phys.\ \textbf{6}, 167 (1965).
\bibitem{AvHa} D.~ben-Avraham and S.~Havlin, \textit{Diffusion and Reactions in Fractals and Disordered Systems} (Cambridge University Press, Cambridge, 2000).
\bibitem{MeKl} R.~Metzler and J.~Klafter, Phys.\ Rep.\ \textbf{339}, 1 (2000).
\bibitem{MJCB} R.~Metzler, J.-H.~Jeon, A.~G.~Cherstvy, and E.~Barkai, Phys.\ Chem.\ Chem.\ Phys.\ \textbf{16}, 24128 (2014).

\bibitem{SBM} J.~H.~P.~Schulz, E.~Barkai, and R.~Metzler, Phys.\ Rev.\ X \textbf{4}, 011028 (2014).

\bibitem{DKDH1} S.~I.~Denisov, M.~Kostur, E.~S.~Denisova, and P.~H\"{a}nggi, Phys.\ Rev.\ E \textbf{75}, 061123 (2007); \textit{ibid.} \textbf{76}, 031101 (2007).

\bibitem{GoLu} C.~Godr\`{e}che and J.~M.~Luck, J.\ Stat.\ Phys.\ \textbf{104}, 489 (2001).

\bibitem{BNO} L.~Beghin, L.~Nieddu, and E.~Orsingher, J.\ Appl.\ Math.\ Stoch.\ Anal.\ \textbf{14}, 11 (2001).
\bibitem{Uch} V.~V.~Uchaikin, Physics\ Uspekhi \textbf{46}, 821 (2003).

\bibitem{DiCr}
    A.~Di~Crescenzo, Adv.\ Appl.\ Prob.\ \textbf{33}, 690 (2001).

\bibitem{Erd}
    A.~Erd\'{e}lyi, ed., \textit{Tables of Integral Transforms}, Bateman Manuscript Pro\-ject, Vol.~1 (McGraw-Hill, New York, 1954), Chap.~5.2.

\bibitem{Fel} W.~Feller, \textit{An Introduction to Probability Theory and its Applications}, Vol.~2 (Wiley, New York, 1971).
\bibitem{Hug} B.~D.~Hughes, \textit{Random Walks and Random Environments}, Vol.~2 (Clarendon Press, Oxford, 1995).

\bibitem{DDK} S.~I.~Denisov, E.~S.~Denisova, and H.~Kantz, Eur.\ Phys.\ J.\ B  \textbf{76}, 1 (2010).

\bibitem{DeK1} S.~I.~Denisov and H.~Kantz, Europhys.\ Lett.\ \textbf{92}, 30001 (2010).
\bibitem{DeK2} S.~I.~Denisov and H.~Kantz, Phys.\ Rev.\ E \textbf{83}, 041132 (2011).
\bibitem{DYBK} S.~I.~Denisov, S.~B.~Yuste, Yu.~S.~Bystrik, H.~Kantz, and K.~Lindenberg, Phys.\ Rev.\ E  \textbf{84}, 061143 (2011).
\bibitem{DBK} S.~I.~Denisov, Yu.~S.~Bystrik, and H.~Kantz, Phys.\ Rev.\ E  \textbf{87}, 022117 (2013).

\bibitem{BGT} N.~H.~Bingham, C.~M.~Goldie, and J.~L.~Teugels, \textit{Regular Variation} (Cambridge University Press, Cambridge, 1987).

\end{thebibliography}
\end{document}